\documentclass[final,5p,times,twocolumn]{elsarticle}

\usepackage[numbers]{natbib}

\usepackage{amsmath,amssymb}

\usepackage{amsthm}
\usepackage{newtxtext,newtxmath}

\usepackage{booktabs}
\usepackage{multirow}
\usepackage{array}
\usepackage{tabularx}
\usepackage{graphicx}
\usepackage{algorithm}
\usepackage{algpseudocode}
\usepackage{lineno}
\usepackage{enumitem}
\usepackage{hyperref}
\usepackage{xcolor}
\usepackage{comment}

\makeatletter
\@fleqnfalse
\makeatother

\newcommand{\err}[2]{\ensuremath{#1\,\scalebox{0.7}{$\pm\,#2$}}}
\newcommand{\errb}[2]{\ensuremath{\mathbf{#1}\,\scalebox{0.7}{$\mathbf{\pm}\,\mathbf{#2}$}}}

\definecolor{mygreen}{HTML}{2E8B57}

\theoremstyle{plain}

\theoremstyle{definition}

\newcommand{\NegEnt}{\mathcal{E}}          % negative-input entropy
\newcommand{\vect}[1]{\mathbf{#1}}
\DeclareMathOperator{\softmax}{softmax}

\journal{Pattern Recognition Letters}

\begin{document}
\let\WriteBookmarks\relax
\def\floatpagepagefraction{1}
\def\textpagefraction{.001}

% ==================================================================
% FRONTMATTER (Required for elsarticle)
% ==================================================================
\begin{frontmatter}

\title{ConCA: Concentration-Aware Channel Attention for Fine-Grained Visual Recognition}

\author[1]{Yu-Sheng Liu}
\ead{611433004@mail.nknu.edu.tw} 

\author[1]{Yu-Chen Tung\corref{cor1}}
\ead{yctung@mail.nknu.edu.tw}
\cortext[cor1]{Corresponding author}

\address[1]{Department of Physics, National Kaohsiung Normal University, Kaohsiung, 824, Taiwan}

\begin{abstract}
Lightweight channel attention mechanisms are widely used in image classification, yet their effectiveness in fine-grained visual recognition (FGVR) remains limited.
Most modules summarize each channel by global average pooling (GAP), which captures activation magnitude but ignores spatial concentration, so channels with different spatial distributions but identical means receive the same descriptor.
We propose Concentration-Aware Channel Attention (ConCA), which pairs the mean with a shift-invariant negative-input entropy (NegEnt), computed via a softmax over the negated activations, forming a dual descriptor that jointly encodes magnitude and concentration.
A depthwise 1-D convolutional multi-layer perceptron (MLP), whose parameter count is linear in the number of channels, maps the pair to a per-channel weight. On six fine-grained benchmarks, ConCA improves over attention-free, SE-Net, and ECA-Net baselines as well as four richer descriptor-based modules under a controlled from-scratch protocol, and it generalizes across eight backbones on iNat2021-mini.
These results indicate that the channel descriptor, 
together with the per-channel gating that maps it to attention weights, 
is an important but underexplored aspect of lightweight channel attention in FGVR.
\end{abstract}


\end{frontmatter}

%==================================================================
\section{Introduction}\label{sec:intro}
%==================================================================

Fine-grained visual recognition (FGVR) distinguishes visually similar subcategories, such as bird species or aircraft models, whose discriminative cues are often confined to small local regions, a beak shape or a wing stripe, rather than the object’s global appearance. Feature representations must therefore preserve the spatial characteristics of discriminative regions while retaining their semantic information.

Channel attention is a widely used lightweight mechanism 
that adaptively recalibrates channel responses. 
Representative modules such as SE-Net~\cite{hu2018squeeze} and ECA-Net~\cite{wang2020eca} differ in how the attention weights are produced, but both summarize each channel by global average pooling (GAP), which captures activation magnitude while discarding the spatial distribution of responses. Localized and diffuse responses with identical means therefore receive the same descriptor, an ambiguity that matters in FGVR, where discriminative evidence is highly localized. Consistent with this, SE-Net and ECA-Net rarely improve on the attention-free baseline in our fine-grained experiments, with only modest gains on iNat2021-mini. This behavior motivates our central question: can a channel descriptor that also encodes spatial concentration, rather than the mean alone, improve lightweight channel attention in FGVR?

We propose Concentration-Aware Channel Attention (ConCA), which describes each channel by two statistics: the mean, reflecting the overall response level, and a negative-input entropy (NegEnt), the entropy of a softmax over the negated activations, which is shift-invariant and measures spatial concentration independently of magnitude. 
A lightweight depthwise 1-D convolutional multi-layer perceptron (MLP) maps this pair to a per-channel attention weight, operating independently on each channel while requiring only a number of parameters linear in the channel count, and ConCA serves as a drop-in module in standard convolutional backbones, where surrounding layers already provide sufficient cross-channel interaction.
Existing channel-attention research has largely focused on designing the descriptor-to-gate mapping. This work instead treats the channel descriptor, the information supplied to that mapping, as the central design axis, deliberately keeping the mapping simple.

The main contributions of this work are as follows:
\begin{itemize}[leftmargin=*,itemsep=2pt,topsep=2pt]
\item \textbf{The channel descriptor as a design axis.} We highlight the channel descriptor used to generate attention weights as an underexplored design axis for lightweight channel attention in FGVR, and show that a mean-only descriptor cannot distinguish channels with identical means but different spatial concentrations.

\item \textbf{Complementary descriptor design.} We propose a dual descriptor combining the channel mean and the NegEnt, capturing both activation magnitude and spatial concentration with minimal parameter overhead, and instantiate it as ConCA, 
which couples this descriptor with per-channel gating.

\item \textbf{Empirical validation.} We evaluate ConCA on six FGVR benchmarks, where it consistently improves over SE-Net, ECA-Net, the attention-free baseline, and four richer descriptor-based modules under a controlled from-scratch protocol, and we confirm cross-architecture generalization across eight backbones on iNat2021-mini.
\end{itemize}

%==================================================================
\section{Related Work}\label{sec:related}
%==================================================================

SE-Net~\cite{hu2018squeeze} introduced the GAP $\to$ fully connected (FC) $\to$ gate paradigm, and 
ECA-Net~\cite{wang2020eca} replaced the FC bottleneck with a shared-kernel 1-D convolution, suggesting that lightweight local channel interaction can be sufficient for effective channel recalibration.
Later modules extend channel attention by enriching either the attention mechanism or the descriptor. CBAM~\cite{woo2018cbam} augments channel attention with a spatial branch and combines GAP with global max pooling (GMP) in its channel descriptor. FcaNet~\cite{qin2021fcanet} interprets GAP as the lowest discrete cosine transform (DCT) frequency and introduces additional frequencies, and SRM~\cite{lee2019srm} uses channel mean and standard deviation.
These designs differ mainly in the descriptor-to-gate mapping or the pooling statistic. However, most descriptors remain first- or second-order summaries and do not explicitly characterize the spatial distribution of activations. 

A related line of work uses entropy as a descriptor. Wan et al.\ \cite{wan2019entropy} weight features by entropy before pooling. Filus and Doma\'nska \cite{filus2023gep} propose a parameter-free Global Entropy Pooling layer and argue analytically that entropy captures distribution shape. CAT~\cite{wu2023cat} fuses GAP, GMP, and an entropy pooler through a shared MLP and couples channel with spatial attention, making it the most closely related design since it also uses an entropy-based descriptor.
Unlike CAT's positive-input entropy (PosEnt) and shared channel-mixing MLP, ConCA pairs the mean with a negative-input entropy and gates each channel independently, differing along both the descriptor and the gate axes. We compare against CAT directly in Section~\ref{sec:exp:fgvr}.
Table~\ref{tab:designspace} compares these modules by descriptor, gating strategy, and parameter cost.
Because CBAM and CAT couple channel with spatial attention, 
the table lists only their channel branches, denoted CBAM-c and CAT-c. 
Along these axes, SRM is the closest counterpart, likewise pairing the mean with a shift-invariant statistic under a per-channel gate at linear cost, but it measures dispersion (std) rather than concentration.

\begin{table}[t]
\caption{Design choices of representative channel-attention modules. 
Shift-inv.\ statistic indicates whether the descriptor includes a shift-invariant spatial statistic, and per-ch. gate indicates whether the attention weight depends only on the channel's own descriptor.
Parameter cost is reported as an order with respect to the channel count $C$ ($k$: ECA kernel size, a small
constant). PosEnt and NegEnt are the softmax entropies of the activations and of their negation, respectively.}\label{tab:designspace}
\centering\footnotesize
\setlength{\tabcolsep}{3pt}
\begin{tabular*}{\columnwidth}{@{\extracolsep{\fill}}lcccl@{}}
\toprule
Method & Descriptor & \shortstack{Shift-inv.\\ statistic} & 
\shortstack{Per-ch.\\gate} & Params \\
\midrule
SE-Net~\cite{hu2018squeeze}   & mean             & $\times$   & $\times$   & $\mathcal{O}(C^2)$ \\
ECA-Net~\cite{wang2020eca}    & mean             & $\times$   & $\times$   & $\mathcal{O}(k)$   \\
SRM~\cite{lee2019srm}         & mean, std        & $\checkmark$ & $\checkmark$ & $\mathcal{O}(C)$   \\
FcaNet~\cite{qin2021fcanet}   & DCT freqs.       & $\times$   & $\times$   & $\mathcal{O}(C^2)$ \\
CBAM-c~\cite{woo2018cbam}     & mean, max        & $\times$   & $\times$   & $\mathcal{O}(C^2)$ \\
CAT-c~\cite{wu2023cat}        & mean, max, PosEnt  & $\checkmark$ & $\times$   & $\mathcal{O}(C^2)$ \\
\textbf{ConCA (Ours)}         & mean, NegEnt & $\checkmark$ & $\checkmark$ & $\mathcal{O}(C)$   \\
\bottomrule
\end{tabular*}
\end{table}%

Because fine-grained categories differ mainly in subtle local cues, global features alone are weakly discriminative~\cite{wei2022fine}. Existing approaches localize discriminative parts~\cite{zhang2014part,branson2014bird}, model high-order feature interactions such as bilinear pooling~\cite{lin2015bilinear}, or learn to attend end-to-end~\cite{zheng2017learning}. Channel attention is a lower-cost alternative, yet GAP-based modules gain little in this setting. ConCA instead operates at the descriptor level, requiring neither extra annotations nor additional feature-encoding modules.

%==================================================================
\section{Method}\label{sec:method}
%==================================================================

\subsection{Motivation and Theory}\label{sec:method:motivation}

Let a feature map at an intermediate layer be
$\vect{X} \in \mathbb{R}^{B \times C \times H \times W}$,
where $B$, $C$, $H$, and $W$ denote the batch size, number of channels,
and spatial dimensions, respectively. Let $N=H\times W$ be the number of
spatial positions, and denote the activation vector of channel $c$ as
$\vect{x}_c=(x_c^{(1)},\dots,x_c^{(N)})\in\mathbb{R}^{N}$.
GAP collapses it to the spatial mean
\begin{equation}\label{eq:mean}
\mu_c = \frac{1}{N}\sum_{i=1}^{N} x_c^{(i)} .
\end{equation}

\paragraph{Ambiguity of the mean descriptor}
As a map $\mathbb{R}^{N}\!\to\!\mathbb{R}$, \eqref{eq:mean} is many-to-one: very different spatial patterns can share the same mean. 
A channel with a single strong activation over a low background and a channel with a spatially uniform response can share the same mean $\mu$, yet GAP maps both to the same value and cannot distinguish them.

\begin{figure}[t]
\centering
\includegraphics[width=0.9\linewidth]{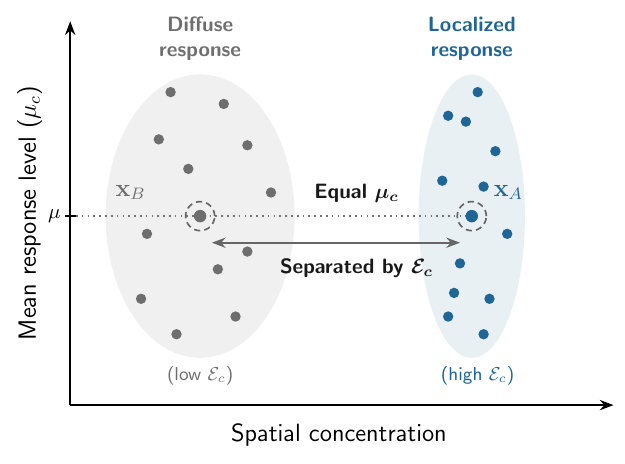}
\caption{Descriptor space (\emph{conceptual illustration}). 
Channels with localized responses and with broadly activated responses overlap along the
mean-response axis but separate along spatial concentration: two
channels of equal mean (circled) receive the same value from a mean-only
descriptor, whereas the pair $(\mu_c,\NegEnt_c)$ distinguishes them, since
 $\NegEnt_c$ decreases as the high-activation region expands and the low-activation background shrinks.}
\label{fig:descspace}
\end{figure}

\paragraph{NegEnt: a complementary concentration descriptor}
To recover the discarded shape information, we apply a softmax to the \emph{negated} activations and take the entropy of the resulting distribution,
\begin{equation}\label{eq:negent}
\NegEnt_c = -\sum_{i=1}^{N} \tilde{p}_i^{(c)} \log \tilde{p}_i^{(c)},~\text{where}~
\tilde{p}_i^{(c)} = \frac{e^{-x_c^{(i)}}}{\sum_{j=1}^{N} e^{-x_c^{(j)}}}.
\end{equation}
Since $\tilde p_i^{(c)}$ is large where the activation is low,
$\NegEnt_c$ measures the spread of this induced softmax distribution and thus reflects the spatial concentration of the response rather than its absolute activation level.
Two properties make it complementary to the mean descriptor. (i)~It is \emph{shift-invariant}: 
a constant offset $\alpha$ cancels in the softmax, $e^{-(x+\alpha)}/\sum_j e^{-(x_j+\alpha)} = e^{-x}/\sum_j e^{-x_j}$, so $\NegEnt_c(\vect{x}_c+\alpha\vect{1}) = \NegEnt_c(\vect{x}_c)$ ($\vect{1}$ the all-ones vector) while the mean shifts by $\alpha$. (ii)~It 
provides \emph{non-redundant spatial-concentration information}. For a localized high-activation pattern over a broad low background (typical in FGVR), the mass of $\tilde{p}^{(c)}$ spreads across the numerous low-activation positions, yielding a high $\NegEnt_c \approx \log N$. As the activated region expands and the low-activation background shrinks, the mass of $\tilde{p}^{(c)}$ concentrates on the fewer remaining low-activation positions, so $\NegEnt_c$ decreases.
The mean and NegEnt therefore capture different aspects of the response, a shift-sensitive activation level and a shift-invariant concentration. The pair $(\mu_c,\NegEnt_c)$ retains the mean while adding a concentration axis, separating channels that GAP alone conflates (Fig.~\ref{fig:descspace}).

\paragraph{Choice of entropy polarity}
The entropy of a softmax can be computed from either the activations
(PosEnt) or their negation (NegEnt), and both are shift-invariant.
CAT~\cite{wu2023cat} adopts PosEnt, whereas ConCA uses NegEnt, and the two differ once paired with the mean. For the localized responses typical of FGVR, $\softmax(\vect{x}_c)$ concentrates on the few dominant activations, so its entropy saturates near zero and PosEnt mainly reflects the magnitude of the strongest responses, information the mean already carries (Section~\ref{sec:exp:redundancy} quantifies this overlap). Negating the input instead spreads mass across the low-activation background, yielding the concentration measure described above. The ablation in Section~\ref{sec:exp:ablation} supports this choice: mean+NegEnt ($52.76$) outperforms mean+PosEnt ($51.83$), which falls below the mean alone ($51.96$).

\subsection{Module Design}\label{sec:method:design}

\begin{figure*}[t]
\centering
\includegraphics[width=0.9\linewidth]{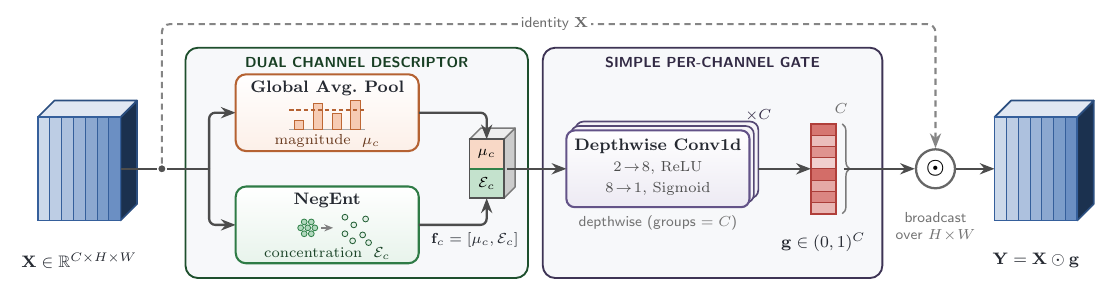}%
\caption{
ConCA overview. Each channel of $\vect{X}$ is summarized by the mean
$\mu_c$ and the NegEnt $\NegEnt_c$. 
The pair $\vect{f}_c=[\mu_c,\NegEnt_c]$ is mapped to a per-channel gate $g_c\in(0,1)$
by a two-layer depthwise 1-D convolutional MLP and applied as
$\vect{Y}=\vect{X}\odot\vect{g}$. The depthwise mapping makes $g_c$ depend
only on channel $c$, meaning that each gate depends only on its own channel descriptor.}
\label{fig:arch}
\end{figure*}

ConCA (Fig.~\ref{fig:arch}) maps an input
$\vect{X}\in\mathbb{R}^{B\times C\times H\times W}$ to
$\vect{Y}=\vect{X}\odot\vect{g}$, where
$\vect{g}=(g_1,\dots,g_C)$ and $g_c\in(0,1)$. 
Each gate is broadcast over the spatial dimensions, recalibrating its channel while preserving the spatial resolution.

\paragraph{Descriptors}
The input is flattened to $\mathbb{R}^{B \times C \times N}$ ($N = H \times W$) and summarized by two per-channel statistics: the mean $\mu_c$ of \eqref{eq:mean} and the NegEnt $\NegEnt_c$ of \eqref{eq:negent}. For numerical stability, $\NegEnt_c$ is evaluated through $\log\softmax(-\vect{x}_c)$, whose internal log-sum-exp avoids forming $e^{-x_c^{(i)}}$ directly. 
Stacking these statistics gives the channel descriptor
$\vect{f}_c=[\mu_c,\NegEnt_c]\in\mathbb{R}^{m}$ with $m=2$.

\paragraph{Depthwise MLP}
The descriptor $\vect{f}_c$ is mapped to a gate by a two-layer depthwise 1-D convolutional MLP (kernel size $1$, groups $= C$), applied identically and independently to every channel:

\begin{align}
\vect{h}_c &= \mathrm{ReLU}\big(\mathrm{Conv1d}^{(1)}_{\mathrm{dw}}(\vect{f}_c)\big), \quad \vect{h}_c \in \mathbb{R}^{4m}, \label{eq:mlp1}\\
g_c &= \mathrm{Sigmoid}\big(\mathrm{Conv1d}^{(2)}_{\mathrm{dw}}(\vect{h}_c)\big), \quad g_c \in (0,1). \label{eq:mlp2}
\end{align}
The first layer expands the $m$ descriptors to a hidden width $4m$ and the second compresses them to the single gate $g_c$, applied as $\vect{Y}_{c,h,w} = g_c\,\vect{X}_{c,h,w}$. 
Because both convolutions are depthwise (groups $= C$), $g_c$ depends only on $(\mu_c,\NegEnt_c)$, with no cross-channel interaction. 
Since the surrounding convolutional layers in standard backbones already mix channels, this isolates the descriptor's effect, spending capacity on the descriptor-to-gate mapping rather than on further cross-channel interaction.
This design assumes dense cross-channel interaction in the backbone, which keeps per-channel statistics meaningful and suits ConCA to standard convolutional backbones. Architectures that structurally restrict such mixing lie outside its intended scope by design (Section~\ref{sec:exp:arch}).

\subsection{Complexity Analysis}\label{sec:method:complexity}

The learnable parameters of ConCA reside entirely in the depthwise MLP.
For $m=2$, the first depthwise layer contributes
$4m^2C+4mC=24C$ parameters (weights and biases), and the second contributes
$4mC+C=9C$, giving
\begin{equation}
P_{\mathrm{ConCA}} = 33C,
\end{equation}
which grows linearly with the number of channels.
For comparison, SE-Net's two fully connected layers require
$P_{\mathrm{SE}}\approx 2C^2/r$ parameters, which grow quadratically with
$C$ ($r=16$), whereas ECA-Net uses a shared 1-D convolution with a small
adaptive kernel 
($P_{\mathrm{ECA}}=k$, a small constant). 
The counts of ConCA and SE-Net are comparable near $C=256$, whereas at $C=2048$, typical of the final stages of ResNet-like convolutional neural networks (CNNs), ConCA uses only about $13\%$ of SE-Net's parameters. 
Both the mean and NegEnt are computed in $\mathcal{O}(CN)$, so NegEnt does not change the asymptotic complexity. 
Table~\ref{tab:cost} reports the measured cost on ResNet-50 using fvcore.FlopCountAnalysis.
ConCA adds negligible parameters ($+0.50$M over the attention-free baseline, against $+2.53$M for SE) and essentially no FLOPs, but its eager-mode latency is higher because NegEnt launches several small reduction and element-wise kernels rather than added arithmetic. 
\texttt{torch.compile} (\texttt{reduce-overhead}) reduces the latency from $452.6$ to $210.8$~ms, indicating that much of the overhead stems from framework execution. Despite nearly identical GFLOPs ($4.11$), the remaining gap reflects memory movement from multiple small kernels rather than additional computation. 
Fusing the softmax and entropy reductions of NegEnt into a single on-chip kernel is a promising direction for reducing this overhead and will be investigated in future work.

\begin{table}[h]
\caption{Whole-model efficiency on ResNet-50 (ImageNet-1K setting:
$224\times224$ input, $1{,}000$ classes) on a single NVIDIA RTX~4090 in PyTorch
(FP32). Latency is per batch of $512$, averaged over $100$ iterations after
$10$ warm-ups. The compiled column uses \texttt{torch.compile} (\texttt{reduce-overhead}).}\label{tab:cost}
\centering\small
\setlength{\tabcolsep}{3pt}
\begin{tabular}{lrrrr}
\toprule
 & & & \multicolumn{2}{c}{Latency (ms)} \\
\cmidrule(l){4-5}
Method & Params (M) & GFLOPs & {eager} & {compiled} \\
\midrule
Baseline & {25.56} & {4.11} & {218.80} & {142.77} \\
SE~\cite{hu2018squeeze} & {28.09} & {4.12} & {255.76} & {178.81} \\
ECA~\cite{wang2020eca} & {25.56} & {4.12} & {255.61} & {178.57} \\
\textbf{ConCA (Ours)} & {26.06} & {4.11} & {452.59} & {210.79} \\
\bottomrule
\end{tabular}
\end{table}

%==================================================================
\section{Experiments}\label{sec:exp}
%==================================================================

\subsection{Setup}\label{sec:exp:setup}

\paragraph{Datasets}
We evaluate ConCA on six FGVR benchmarks:
CUB\allowbreak-200\allowbreak-2011~\cite{wah2011caltech}, 
FGVC\allowbreak-Aircraft~\cite{maji2013fine},
Stanford Cars~\cite{krause20133d}, 
Stanford Dogs~\cite{khosla2011novel},
NABirds~\cite{vanhorn2015building}, 
and Oxford-IIIT Pets~\cite{parkhi2012cats}.
These datasets contain 37--555 sub-categories 
with strong intra-class variation and subtle inter-class differences.
We additionally evaluate cross-architecture generalization on the
large-scale iNat2021-mini~\cite{vanhorn2018inaturalist,vanhorn2021benchmarking},
which contains 10,000 species with approximately 50 images per class.
CUB, Cars, Dogs, NABirds, and Pets use a fixed 70/15/15 split (seed 42), 
and Aircraft uses its official split. CUB and Aircraft use the provided
bounding boxes. For iNat2021-mini, we train on 45 images per class,
validate on 5 images per class, and evaluate on the official validation set.

\paragraph{Training protocol}
Fine-grained recognition and channel-attention research follow two different conventions, and because ConCA sits at their intersection we evaluate under both. Following the standard FGVR protocol, we fine-tune an ImageNet~\cite{russakovsky2015imagenet}-pretrained ResNet-50~\cite{he2016deep} on three benchmarks (Section~\ref{sec:exp:pretrain}). Following the common channel-attention protocol, we train ResNet-18 from scratch on all six benchmarks, which isolates the module's contribution from dominant pretrained features and confirms that the baseline behavior is not an artifact of either regime (Section~\ref{sec:exp:fgvr}). Cross-architecture generalization is then evaluated on iNat2021-mini using from-scratch training (Section~\ref{sec:exp:arch}).

\paragraph{Compared methods}
We compare ConCA with the attention-free \textbf{Baseline}, the GAP-based \textbf{SE-Net}~\cite{hu2018squeeze} and \textbf{ECA-Net}~\cite{wang2020eca}, and four richer descriptor-based modules: \textbf{SRM}~\cite{lee2019srm} (mean and standard deviation), \textbf{FcaNet}~\cite{qin2021fcanet} (multi-frequency DCT), \textbf{CBAM-c}~\cite{woo2018cbam} (GAP $+$ GMP), and \textbf{CAT-c}~\cite{wu2023cat} (GAP $+$ GMP $+$ PosEnt). CAT-c is the most direct comparison, as it also incorporates an entropy descriptor. For CBAM and CAT we evaluate only their channel branches, denoted CBAM-c and CAT-c, so that the comparison stays on the channel-descriptor axis. Their spatial branches are orthogonal to descriptor design and out of scope here. All methods use their recommended configurations, with hyper-parameters shared unless a method-specific default is required.

\paragraph{Hyper-parameters}
For the six FGVR benchmarks, all methods are trained with ResNet-18 using
AdamW~\cite{loshchilov2019decoupled} ($\mathrm{lr}=10^{-3}$,
weight decay $10^{-2}$), cosine annealing learning rate scheduling
($T_{\max}=90$), batch size 128, and 90 epochs.
Training uses cross-entropy loss, mixed precision, gradient clipping
(1.0), and standard ImageNet normalization.
Data augmentation includes RandomResizedCrop(224), horizontal flipping,
and ColorJitter. 
CenterCrop(224) is used during testing.
For all experiments, ConCA uses hidden width $4m=8$ and is inserted at the
same position as SE-Net.
We report mean $\pm$ std over five seeds (43--47) for FGVR benchmarks and
three seeds (43--45) for iNat2021-mini.
Model selection uses validation accuracy, and final results are reported as
test top-1 accuracy.

\subsection{Validation in the Pretrained Setting}\label{sec:exp:pretrain}

FGVR models are commonly initialized from ImageNet-pretrained backbones, so we first ask whether GAP-based channel attention behaves differently under this standard transfer-learning setting. Table~\ref{tab:pretrain} reports fine-tuning of an ImageNet-pretrained ResNet-50 on CUB-200, Aircraft, and Cars. 
Because SE, ECA, and ConCA are newly inserted layers absent from the pretrained weights, their final gating layers are initialized so that each gate (the Sigmoid output) is close to one, an approximate identity mapping that leaves the pretrained features unchanged at the start of fine-tuning.
All methods perform similarly, with average accuracies differing by only $0.65$ points. 
Since every module starts from a near-identity mapping and the pretrained features are already highly discriminative, the room for lightweight channel recalibration is limited. 
We therefore focus on the from-scratch setting (Section~\ref{sec:exp:fgvr}), 
where the descriptor’s contribution is more directly observable.

\begin{table}[h]
\caption{Top-1 accuracy (\%, mean $\pm$ std over 5 seeds) under the
standard FGVR transfer-learning setting. An ImageNet-pretrained ResNet-50
is fine-tuned on CUB-200, Aircraft, and Cars.
The Avg column reports the unweighted mean of the three dataset means, with uncertainty computed by propagating the per-dataset standard deviations assuming independent runs across datasets.}\label{tab:pretrain}
\centering\footnotesize
\setlength{\tabcolsep}{2pt}
\begin{tabular*}{\columnwidth}{@{\extracolsep{\fill}}lcccc@{}}
\toprule
Method & CUB & Air. & Cars & Avg \\
\midrule
Baseline & 70.76 $\pm$ 1.09 & 48.26 $\pm$ 0.59 & 81.83 $\pm$ 0.69 & 66.95 $\pm$ 0.47 \\
SE \cite{hu2018squeeze}  & 70.27 $\pm$ 0.82 & 48.56 $\pm$ 0.18 & 81.28 $\pm$ 0.38 & 66.70 $\pm$ 0.31 \\
ECA \cite{wang2020eca} & 70.54 $\pm$ 0.39 & 48.38 $\pm$ 1.00 & 81.70 $\pm$ 0.85 & 66.87 $\pm$ 0.46 \\
\textbf{ConCA (Ours)} & \textbf{71.07 $\pm$ 0.79} & \textbf{48.93 $\pm$ 0.50} & \textbf{82.05 $\pm$ 0.70} & \textbf{67.35 $\pm$ 0.39} \\
\bottomrule
\end{tabular*}
\end{table}

\subsection{Fine-Grained Benchmark Comparison}\label{sec:exp:fgvr}

Table~\ref{tab:fgvr} reports the from-scratch comparison using ResNet-18,
which isolates the contribution of the channel-attention module itself.
Both GAP-only modules yield lower accuracy than the baseline across all six datasets, with average drops of 5.34 and 2.34 percentage points for SE and ECA, respectively.
ConCA exceeds the baseline on all six datasets ($+1.59$ on average) 
and achieves the highest mean accuracy among the evaluated channel-attention modules on every benchmark. 
One-sided Welch's $t$-tests over the five seeds (marked in Table~\ref{tab:fgvr}) show that ConCA significantly improves ($p<0.05$) over SE on all six datasets and over ECA on four. 
Over the strong attention-free baseline, the gains are significant on CUB-200, NABirds, and Pets and lie within run-to-run variation on Aircraft, Cars, and Dogs, where the margins are small.
ConCA also has lower run-to-run variance than the Baseline, SE, and ECA on every dataset except Cars, where variance is large for all methods yet ConCA still attains the highest mean accuracy, $1.67$ points above the baseline.

The richer descriptor-based modules also fail to consistently improve over the baseline. CAT-c, despite using an entropy descriptor, trails ConCA by 6.88 points, while SRM, the strongest competing module, remains 3.79 points lower on average. ConCA is the only method that improves over the baseline on all six benchmarks.
These comparisons suggest that effective channel attention in fine-grained recognition benefits from pairing a non-redundant descriptor with per-channel gating, rather than from entropy or a richer descriptor in isolation.

\begin{table*}[h]
\caption{Top-1 accuracy ($\%$, mean $\pm$ std over 5 seeds) on six fine-grained benchmarks using ResNet-18 trained from scratch. 
The best result in each column is shown in bold. 
\dag/\ddag/\S: $p<0.05$ (one-sided Welch's $t$-test, 5 seeds) for ConCA vs.\ Baseline/SE/ECA, respectively.
Avg follows the averaging convention of Table~\ref{tab:pretrain}.}
\label{tab:fgvr}
\centering\footnotesize
\setlength{\tabcolsep}{4pt}
\begin{tabular}{lccccccc}
\toprule
Method & CUB-200 & Aircraft & Cars & Dogs & NABirds & Pets & Avg \\
\midrule
Baseline & 58.98 $\pm$ 0.96 & 41.57 $\pm$ 0.76 & 53.80 $\pm$ 4.37 & 60.79 $\pm$ 0.56 & 69.29 $\pm$ 0.61 & 55.94 $\pm$ 1.95 & {56.73 $\pm$ 0.83} \\
SE \cite{hu2018squeeze}  & 53.85 $\pm$ 0.96 & 38.06 $\pm$ 1.18 & 41.11 $\pm$ 1.01 & 56.21 $\pm$ 0.60 & 67.13 $\pm$ 0.32 & 51.99 $\pm$ 2.51 & {51.39 $\pm$ 0.53} \\
ECA \cite{wang2020eca} & 55.39 $\pm$ 1.38 & 40.15 $\pm$ 1.43 & 50.46 $\pm$ 3.44 & 58.24 $\pm$ 0.80 & 67.10 $\pm$ 0.40 & 55.00 $\pm$ 2.79 & {54.39 $\pm$ 0.82} \\
SRM \cite{lee2019srm} & 57.88 $\pm$ 1.04 & 39.11 $\pm$ 0.91 & 48.07 $\pm$ 0.58 & 57.95 $\pm$ 0.80 & 67.99 $\pm$ 0.23 & 56.19 $\pm$ 1.76 & {54.53 $\pm$ 0.41} \\
FcaNet \cite{qin2021fcanet} & 55.83 $\pm$ 0.25 & 40.78 $\pm$ 0.55 & 46.10 $\pm$ 1.79 & 58.18 $\pm$ 0.92 & 67.58 $\pm$ 0.26 & 55.58 $\pm$ 1.99 & {54.01 $\pm$ 0.48} \\
CBAM-c \cite{woo2018cbam} & 54.91 $\pm$ 1.06 & 40.79 $\pm$ 1.25 & 44.13 $\pm$ 2.65 & 58.39 $\pm$ 0.74 & 66.46 $\pm$ 0.52 & 57.14 $\pm$ 2.76 & {53.64 $\pm$ 0.71} \\
CAT-c \cite{wu2023cat} & 52.04 $\pm$ 0.80 & 39.20 $\pm$ 0.75 & 41.08 $\pm$ 1.47 & 56.16 $\pm$ 0.80 & 63.11 $\pm$ 0.20 & 57.03 $\pm$ 2.50 & {51.44 $\pm$ 0.53} \\
\textbf{ConCA (Ours)} & \textbf{61.15 $\pm$ 0.57}{\textsuperscript{\dag\ddag\S}} & \textbf{41.66 $\pm$ 0.43}{\textsuperscript{\ddag}} & \textbf{55.47 $\pm$ 3.87}{\textsuperscript{\ddag}} & \textbf{61.40 $\pm$ 0.43}{\textsuperscript{\ddag\S}} & \textbf{71.15 $\pm$ 0.22}{\textsuperscript{\dag\ddag\S}} & \textbf{59.09 $\pm$ 1.83}{\textsuperscript{\dag\ddag\S}} & {\textbf{58.32 $\pm$ 0.73}} \\
\bottomrule
\end{tabular}
\end{table*}

\subsection{Cross-Architecture Generalization}\label{sec:exp:arch}

For the cross-architecture study we retain SE and ECA, the most widely adopted baselines, since the richer modules in Table~\ref{tab:fgvr} gave no consistent advantage over ECA in Section~\ref{sec:exp:fgvr}.
Table~\ref{tab:inat} reports results on eight architectures spanning the
ResNet~\cite{he2016deep},
Inception~\cite{szegedy2016rethinking,szegedy2017inception},
DenseNet~\cite{huang2017densely}, and
HRNet~\cite{wang2020deep} families.
These eight backbones are built predominantly from standard convolutions and therefore satisfy ConCA's assumption of dense cross-channel interaction, providing a fair testbed for its cross-architecture generalization.
ConCA achieves the highest accuracy on all eight architectures.
The gains are particularly pronounced on ResNet-50
($+1.85$ percentage points over the second-best method) and
Inception-v4 ($+1.07$), while smaller but consistent improvements are
observed on the remaining architectures.
Unlike on the small FGVR datasets, SE and ECA often improve over the baseline on iNat2021-mini. However, the effect remains architecture dependent, as both fall below the baseline on both DenseNet variants.

A one-sided Welch's $t$-test ($\alpha=0.05$) indicates that ConCA
significantly outperforms the baseline on seven of the eight
architectures, SE on six, and ECA on seven.
The remaining comparisons do not reach statistical significance with
three training seeds.
The dual descriptor's benefit thus generalizes across backbones with substantially different connectivity and feature-aggregation strategies.

\begin{comment}
\begin{table*}[h]
\caption{Top-1 accuracy (\%, mean $\pm$ std over 3 seeds) on iNat2021-mini across eight architectures. $\Delta$ is ConCA minus the second-best method, computed from the reported means.}\label{tab:inat-2col}
\centering\small
\begin{tabular}{lccccc}
\toprule
Architecture & Baseline & SE & ECA & \textbf{ConCA} & $\Delta$ \\
\midrule
ResNet-18     & 43.64 $\pm$ 0.28 & 43.62 $\pm$ 0.19 & 44.47 $\pm$ 0.12 & \textbf{45.10 $\pm$ 0.27} & $+$0.63 \\
ResNet-50     & 48.63 $\pm$ 0.13 & 50.43 $\pm$ 0.88 & 49.66 $\pm$ 1.38 & \textbf{52.28 $\pm$ 0.79} & $+$1.85 \\
ResNet-101    & 48.84 $\pm$ 0.45 & 50.31 $\pm$ 0.74 & 50.72 $\pm$ 0.28 & \textbf{51.21 $\pm$ 0.26} & $+$0.49 \\
Inception-v3  & 56.91 $\pm$ 0.21 & 58.67 $\pm$ 0.08 & 57.33 $\pm$ 0.22 & \textbf{59.01 $\pm$ 0.21} & $+$0.34 \\
Inception-v4  & 58.76 $\pm$ 0.05 & 56.18 $\pm$ 1.31 & 59.88 $\pm$ 0.14 & \textbf{60.95 $\pm$ 0.11} & $+$1.07 \\
DenseNet-121  & 52.76 $\pm$ 0.08 & 50.16 $\pm$ 0.22 & 47.76 $\pm$ 1.11 & \textbf{53.55 $\pm$ 0.17} & $+$0.79 \\
DenseNet-201  & 55.13 $\pm$ 0.12 & 54.22 $\pm$ 0.30 & 54.43 $\pm$ 0.66 & \textbf{55.32 $\pm$ 0.41} & $+$0.19 \\
HRNet-W18-C   & 49.04 $\pm$ 0.91 & 50.40 $\pm$ 0.21 & 48.96 $\pm$ 0.24 & \textbf{50.59 $\pm$ 0.24} & $+$0.19 \\
\bottomrule
\end{tabular}
\end{table*}
\end{comment}

\begin{table}[h]
\caption{Top-1 accuracy (\%, mean $\pm$ std over 3 seeds) on iNat2021-mini across eight architectures. \textbf{ConCA} attains the best accuracy on every architecture.}\label{tab:inat}
\centering\footnotesize
\setlength{\tabcolsep}{3pt}
\begin{tabular*}{\columnwidth}{@{\extracolsep{\fill}}lcccc@{}}
\toprule
Architecture & Baseline & SE & ECA & \textbf{ConCA} \\
\midrule
ResNet-18     & 43.64\,$\pm$\,0.28 & 43.62\,$\pm$\,0.19 & 44.47\,$\pm$\,0.12 & \textbf{45.10\,$\pm$\,0.27} \\
ResNet-50     & 48.63\,$\pm$\,0.13 & 50.43\,$\pm$\,0.88 & 49.66\,$\pm$\,1.38 & \textbf{52.28\,$\pm$\,0.79} \\
ResNet-101    & 48.84\,$\pm$\,0.45 & 50.31\,$\pm$\,0.74 & 50.72\,$\pm$\,0.28 & \textbf{51.21\,$\pm$\,0.26} \\
Inception-v3  & 56.91\,$\pm$\,0.21 & 58.67\,$\pm$\,0.08 & 57.33\,$\pm$\,0.22 & \textbf{59.01\,$\pm$\,0.21} \\
Inception-v4  & 58.76\,$\pm$\,0.05 & 56.18\,$\pm$\,1.31 & 59.88\,$\pm$\,0.14 & \textbf{60.95\,$\pm$\,0.11} \\
DenseNet-121  & 52.76\,$\pm$\,0.08 & 50.16\,$\pm$\,0.22 & 47.76\,$\pm$\,1.11 & \textbf{53.55\,$\pm$\,0.17} \\
DenseNet-201  & 55.13\,$\pm$\,0.12 & 54.22\,$\pm$\,0.30 & 54.43\,$\pm$\,0.66 & \textbf{55.32\,$\pm$\,0.41} \\
HRNet-W18-C   & 49.04\,$\pm$\,0.91 & 50.40\,$\pm$\,0.21 & 48.96\,$\pm$\,0.24 & \textbf{50.59\,$\pm$\,0.24} \\
\bottomrule
\end{tabular*}

\medskip
\footnotesize\textit{iNat schedules (all AdamW + Cosine\allowbreak Annealing\allowbreak LR, 90 epochs unless noted). Abbreviations: bs (batch size), lr (learning rate), wd (weight decay), and $+$w (a 5-epoch linear warm-up from 10\% of the base lr).}
ResNet-18: bs 256, lr $10^{-3}$, wd $10^{-2}$. 
ResNet-50: bs 128, lr $5 \times 10^{-4}$, wd $10^{-2}$. 
ResNet-101: bs 128, lr $10^{-3}$, wd $10^{-4}$. 
Inception-v3: bs 256, lr $10^{-3}$, wd $10^{-4}$, $+$w. 
Inception-v4: bs 128, lr $10^{-3}$, wd $10^{-4}$, $+$w. 
DenseNet-121: bs 256, lr $10^{-3}$, wd $10^{-2}$, $+$w. 
DenseNet-201: bs 128, lr $10^{-3}$, wd $0.05$, $+$w. 
HRNet-W18-C: bs 256, lr $10^{-3}$, wd $10^{-4}$, 120 epochs, $+$w. 
All use ImageNet normalization, RandomResizedCrop(224), horizontal flip, RandomRotation(15) and ColorJitter(0.2, 0.2, 0.2, 0.1).
\end{table}

\subsection{Descriptor Ablation}\label{sec:exp:ablation}

We evaluate descriptor combinations on CUB-200, Aircraft, and Cars over five seeds. The candidates comprise \emph{level descriptors} (mean, max, min), which describe activation magnitude, and \emph{shape descriptors} (std, skew, PosEnt, NegEnt), which characterize the spatial distribution. Preliminary experiments showed little benefit from min, so it appears only in the all-7 combination.

Table~\ref{tab:ablation} indicates that combining complementary descriptor
types outperforms any single statistic: the best single descriptor, mean
($51.96\%$), rises to $52.76\%$ when paired with NegEnt, the highest average
of all combinations. Among shape complements to the mean, NegEnt outperforms
std ($52.45\%$), PosEnt ($51.83\%$), and skew ($50.62\%$). 
PosEnt and NegEnt perform almost identically in isolation ($50.93\%$ vs.\ $50.94\%$), yet only NegEnt gains substantially with the mean, a redundancy effect that Section~\ref{sec:exp:redundancy} quantifies directly.
Skew is unreliable when estimated from
only $N=H\times W$ spatial samples.
Adding more descriptors does not help: mean+max+NegEnt and the all-7 combination fall below mean+NegEnt, the latter sharply ($48.49\%$), suggesting that descriptor quality matters more than quantity. A one-sided Welch's $t$-test on Avg indicates that ConCA significantly outperforms several alternatives, including NegEnt alone ($p=0.018$), std alone ($p=0.043$), mean+skew ($p=0.010$), max+NegEnt ($p=0.026$), and all 7 ($p<0.001$), while the remaining comparisons do not reach significance with five seeds.

\begin{table}[h]
\caption{Descriptor-combination ablation: top-1 accuracy (\%, mean $\pm$ std over 5 seeds) and the across-dataset average (Avg). Single-descriptor rows feed only that descriptor to the depthwise MLP (hidden width $4m$). Avg follows the averaging convention of Table~\ref{tab:pretrain}.}
\label{tab:ablation}
\centering
\footnotesize
\setlength{\tabcolsep}{2pt}
\renewcommand{\arraystretch}{1.0}
\begin{tabular*}{\columnwidth}{@{\extracolsep{\fill}}lcccc@{}}
\toprule
Descriptor & CUB & Air. & Cars & Avg \\
\midrule
mean                  & \err{60.06}{0.81} & \err{41.23}{0.95} & \err{54.59}{1.57} & {\err{51.96}{0.67}} \\
PosEnt                  & \err{58.54}{0.97} & \err{41.03}{0.41} & \err{53.23}{1.31} & {\err{50.93}{0.56}} \\
NegEnt                  & \err{58.43}{0.65} & \err{40.87}{0.78} & \err{53.52}{2.08} & {\err{50.94}{0.77}} \\
max                   & \err{59.46}{0.75} & \err{41.54}{1.28} & \err{52.99}{1.62} & {\err{51.33}{0.73}} \\
std                   & \err{59.81}{0.22} & \err{41.02}{1.10} & \err{53.26}{1.65} & {\err{51.36}{0.67}} \\
mean + std            & \err{61.14}{1.07} & \err{40.91}{0.87} & \err{55.31}{1.75} & {\err{52.45}{0.74}} \\
mean + skew           & \err{58.37}{0.67} & \err{41.31}{0.39} & \err{52.19}{1.40} & {\err{50.62}{0.53}} \\
mean + max            & \err{61.52}{0.60} & \err{41.15}{0.43} & \err{55.24}{0.93} & {\err{52.64}{0.40}} \\
max + NegEnt            & \err{59.89}{0.66} & \err{40.38}{0.59} & \err{52.86}{2.24} & {\err{51.04}{0.80}} \\
std + NegEnt            & \err{60.73}{0.97} & \err{41.02}{0.68} & \err{53.96}{0.96} & {\err{51.90}{0.51}} \\
mean + PosEnt           & \err{60.34}{0.51} & \err{41.09}{0.79} & \err{54.06}{3.21} & {\err{51.83}{1.11}} \\
\textbf{mean + NegEnt}  & \errb{61.15}{0.57} & \errb{41.66}{0.43} & \errb{55.47}{3.87} & {\errb{52.76}{1.31}} \\
mean + PosEnt + NegEnt  & \err{60.43}{0.56} & \err{41.29}{0.92} & \err{54.01}{1.54} & {\err{51.91}{0.63}} \\
mean + max + PosEnt     & \err{61.02}{1.05} & \err{41.49}{1.41} & \err{52.92}{0.87} & {\err{51.81}{0.65}} \\
mean + max + NegEnt     & \err{61.78}{0.94} & \err{41.95}{1.26} & \err{52.75}{1.45} & {\err{52.16}{0.71}} \\
all 7                 & \err{56.90}{1.12} & \err{41.03}{1.17} & \err{47.54}{0.61} & {\err{48.49}{0.58}} \\
\bottomrule
\end{tabular*}
\end{table}

\paragraph{Descriptor versus gate}
The ablation above holds the per-channel gate fixed, isolating the descriptor's contribution \emph{given} that gate. Table~\ref{tab:gatedesc} instead crosses the two descriptors (mean, mean+NegEnt) with two gating strategies (full cross-channel interaction as in SE, and the per-channel gate). Per-channel gating alone already exceeds the attention-free baseline, whereas full interaction falls well below it, and NegEnt adds a further gain on top of the per-channel gate.

\begin{table}[h]
\caption{Gate\,$\times$\,descriptor study (top-1 \%, Avg over CUB-200/Aircraft/Cars, 5 seeds). Avg follows the averaging convention of Table~\ref{tab:pretrain}. Full-interaction and per-channel gates use the same hidden width and training schedule, differing only in cross-channel interaction.}
\label{tab:gatedesc}
\centering\footnotesize
\setlength{\tabcolsep}{3pt}
\begin{tabular*}{\columnwidth}{@{\extracolsep{\fill}}llc@{}}
\toprule
{Gate} & {Descriptor} & {Avg} \\
\midrule
{Baseline (none)}     & {--}                 & {51.45 $\pm$ 1.51} \\
{Full interaction}    & {mean}               & {44.34 $\pm$ 0.61} \\
{Full interaction}    & {mean + NegEnt}      & {44.78 $\pm$ 0.55} \\
{Per-channel}         & {mean}               & {51.96 $\pm$ 0.67} \\
{\textbf{Per-channel}} & {\textbf{mean + NegEnt (ConCA)}} & {\textbf{52.76 $\pm$ 1.31}} \\
\bottomrule
\end{tabular*}
\end{table}

\subsection{Descriptor Redundancy Analysis}\label{sec:exp:redundancy}

\begin{figure}[t]
\centering
\includegraphics[width=0.90\linewidth]{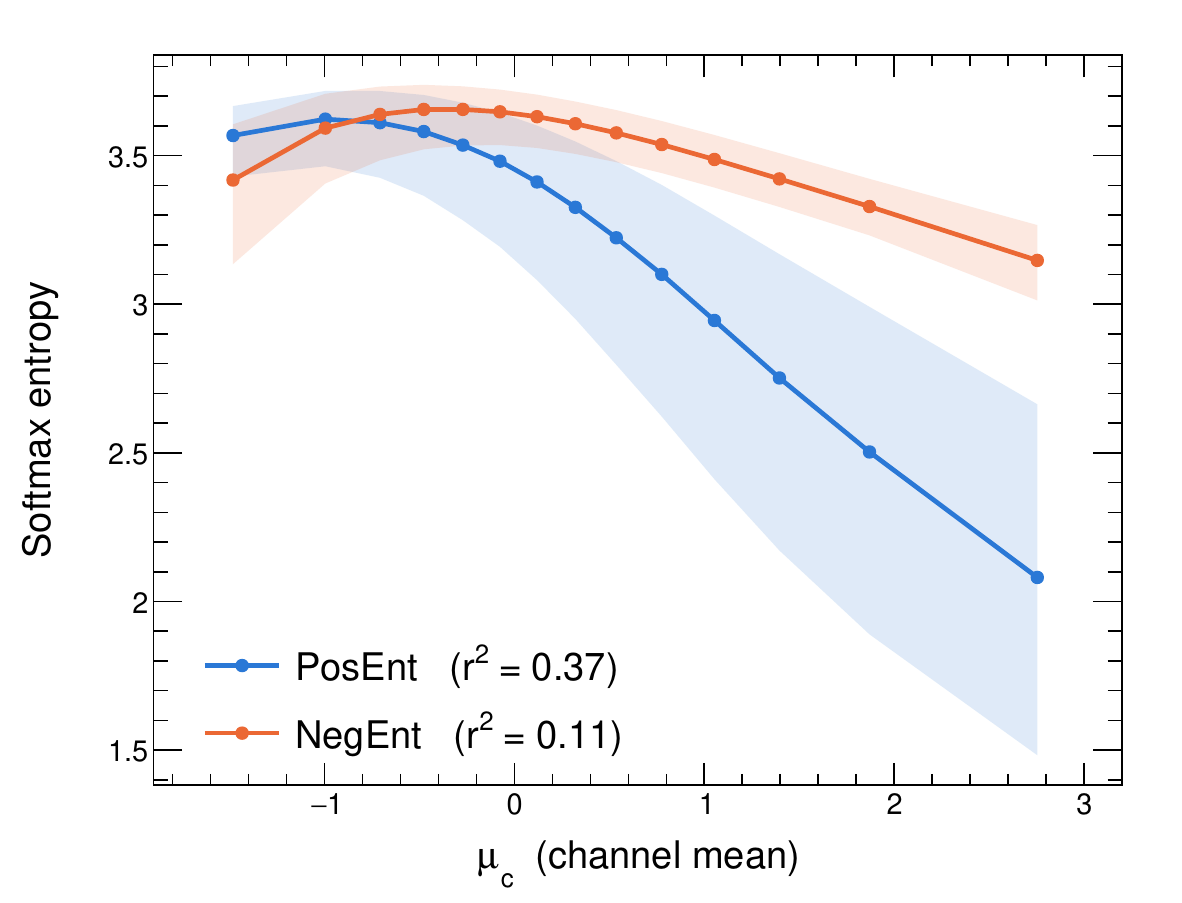}% 
\caption{Redundancy of the entropy descriptors with the channel mean,
measured on the attention-free baseline ResNet-18 (CUB-200 test split, five
seeds pooled, $4.53\times10^{6}$ channel samples). Each curve traces the
per-bin fitted center of an entropy descriptor as a function of the channel
mean $\mu_c$, and the shaded band shows the fitted $\pm1\sigma$ width within
each $\mu_c$ bin. 
PosEnt varies systematically with $\mu_c$ ($r^2=0.37$), whereas NegEnt stays largely flat and is only weakly determined by $\mu_c$ ($r^2=0.11$),
indicating that NegEnt carries channel-level structure not encoded by the
mean.}
\label{fig:redundancy}
\end{figure}

A natural objection is that entropy, as a concentration measure, might merely restate the magnitude already summarized by the mean. We test this directly, without reference to accuracy, by measuring how much of each entropy descriptor is fixed by the channel mean. Using the five attention-free baseline ResNet-18 models of Table~\ref{tab:fgvr}, we extract features at the layer-4 ConCA insertion point on the CUB-200 test split, pool $4.53\times10^{6}$ channel samples, and regress each descriptor on $\mu_c$, reporting the coefficient of determination $r^2$.
As Figure~\ref{fig:redundancy} shows, PosEnt varies systematically with $\mu_c$, so the mean accounts for a sizable share of its variation ($r^2=0.37$), whereas NegEnt stays largely flat while retaining spread at each $\mu_c$, leaving most of its variation unexplained ($r^2=0.11$).
PosEnt thus appears substantially more redundant with the mean than NegEnt, while NegEnt carries spatial structure that the mean does not.

This accounts for the ablation independently of accuracy: because PosEnt is largely co-determined with the mean, adding it helps little (mean+PosEnt $51.83\%$ vs.\ mean $51.96\%$), whereas the near mean-independent NegEnt supplies a complementary axis and yields the best pair ($52.76\%$). 

\section{Conclusion}\label{sec:conclusion}
%==================================================================

We identified the channel descriptor, together with the per-channel gating that maps it to attention weights, as an underexplored design axis in lightweight channel attention for fine-grained recognition. ConCA pairs the shift-sensitive mean with the shift-invariant NegEnt under a depthwise per-channel gate, and it consistently outperforms existing channel-attention modules across six benchmarks and eight CNN architectures at minimal parameter cost, with the gain attributed jointly to the gating and the non-redundant descriptor rather than to entropy alone. ConCA assumes that channels carry meaningful spatial structure and, by design, targets dense-convolution backbones rather than strongly grouped architectures. 
Future work will extend descriptor design to alternative concentration measures, gating strategies, and transformer architectures.

\section*{CRediT authorship contribution statement}
\textbf{Yu-Sheng Liu:} Writing – original draft, Methodology, Software, Investigation.
\textbf{Yu-Chen Tung:} Writing -- review \& editing, Conceptualization, Supervision, Funding acquisition.
 
\section*{Declaration of competing interest}
The authors declare no competing interests.
 
\section*{Data availability}
All datasets used in this study are publicly available.

\section*{Acknowledgements}
The authors acknowledge the computational resources provided through the support of the National Science and Technology Council (NSTC), Taiwan, under Grant No. 114-2112-M-017-001.


\begin{thebibliography}{99}

\bibitem{hu2018squeeze}
J.~Hu, L.~Shen, G.~Sun,
Squeeze-and-excitation networks,
in: Proc.\ IEEE Conf.\ on Computer Vision and Pattern Recognition (CVPR),
2018, pp.\ 7132--7141.

\bibitem{wang2020eca}
Q.~Wang, B.~Wu, P.~Zhu, P.~Li, W.~Zuo, Q.~Hu,
ECA-Net: Efficient channel attention for deep convolutional neural
networks,
in: Proc.\ IEEE/CVF Conf.\ on Computer Vision and Pattern Recognition
(CVPR), 2020, pp.\ 11534--11542.

\bibitem{woo2018cbam}
S.~Woo, J.~Park, J.-Y.~Lee, I.~S.~Kweon,
CBAM: Convolutional block attention module,
in: Proc.\ European Conf.\ on Computer Vision (ECCV), 2018, pp.\ 3--19.

\bibitem{qin2021fcanet}
Z.~Qin, P.~Zhang, F.~Wu, X.~Li,
FcaNet: Frequency channel attention networks,
in: Proc.\ IEEE/CVF Int.\ Conf.\ on Computer Vision (ICCV), 2021,
pp.\ 763--772.

\bibitem{lee2019srm}
H.~Lee, H.~Kim, H.~Nam,
SRM: A style-based recalibration module for convolutional neural networks,
in: Proc.\ IEEE/CVF Conf.\ on Computer Vision and Pattern Recognition
(CVPR), 2019, pp.\ 1854--1863.

\bibitem{wan2019entropy}
W.~Wan, J.~Chen, T.~Li, Y.~Huang, J.~Tian, C.~Yu, Y.~Xue,
Information entropy based feature pooling for convolutional neural
networks,
in: Proc.\ IEEE/CVF Int.\ Conf.\ on Computer Vision (ICCV), 2019,
pp.\ 4311--4320.

\bibitem{filus2023gep}
K.~Filus, J.~Doma\'nska,
Global entropy pooling layer for convolutional neural networks,
Neurocomputing 555 (2023) 126615.

\bibitem{wu2023cat}
Z.~Wu, M.~Wang, W.~Sun, Y.~Li, T.~Xu, F.~Wang, K.~Huang,
CAT: Learning to collaborate channel and spatial attention from
multi-information fusion,
IET Comput.\ Vis.\ 17~(3) (2023) 309--318.

\bibitem{wei2022fine}
X.-S.~Wei, Y.-Z.~Song, O.~Mac Aodha, J.~Wu, Y.~Peng, J.~Tang, J.~Yang,
Fine-grained image analysis with deep learning: A survey,
IEEE Trans.\ Pattern Anal.\ Mach.\ Intell.\ 44~(12) (2022) 9609--9630.

\bibitem{zhang2014part}
N.~Zhang, J.~Donahue, R.~Girshick, T.~Darrell,
Part-based R-CNNs for fine-grained category detection,
in: Proc.\ European Conf.\ on Computer Vision (ECCV), 2014, pp.\ 834--849.

\bibitem{branson2014bird}
S.~Branson, G.~Van Horn, S.~Belongie, P.~Perona,
Bird species categorization using pose normalized deep convolutional nets,
in: Int.\ Conf.\ on Learning Representations (ICLR), 2014.

\bibitem{lin2015bilinear}
T.-Y.~Lin, A.~RoyChowdhury, S.~Maji,
Bilinear CNN models for fine-grained visual recognition,
in: Proc.\ IEEE Int.\ Conf.\ on Computer Vision (ICCV), 2015,
pp.\ 1449--1457.

\bibitem{zheng2017learning}
H.~Zheng, J.~Fu, T.~Mei, J.~Luo,
Learning multi-attention convolutional neural network for fine-grained
image recognition,
in: Proc.\ IEEE Int.\ Conf.\ on Computer Vision (ICCV), 2017,
pp.\ 5209--5217.

\bibitem{wah2011caltech}
C.~Wah, S.~Branson, P.~Welinder, P.~Perona, S.~Belongie,
The Caltech-UCSD Birds-200-2011 dataset,
Tech.\ Rep., California Institute of Technology, 2011.

\bibitem{maji2013fine}
S.~Maji, E.~Rahtu, J.~Kannala, M.~Blaschko,
Fine-grained visual classification of aircraft,
arXiv:1306.5151, 2013.

\bibitem{krause20133d}
J.~Krause, M.~Stark, J.~Deng, L.~Fei-Fei,
3D object representations for fine-grained categorization,
in: Proc.\ IEEE Int.\ Conf.\ on Computer Vision Workshops (ICCVW), 2013,
pp.\ 554--561.

\bibitem{khosla2011novel}
A.~Khosla, N.~Jayadevaprakash, B.~Yao, F.-F.~Li,
Novel dataset for fine-grained image categorization: Stanford Dogs,
in: CVPR Workshop on Fine-Grained Visual Categorization (FGVC), 2011.

\bibitem{vanhorn2015building}
G.~Van Horn, S.~Branson, R.~Farrell, S.~Haber, J.~Barry, P.~Ipeirotis,
P.~Perona, S.~Belongie,
Building a bird recognition app and large scale dataset with citizen
scientists,
in: Proc.\ IEEE Conf.\ on Computer Vision and Pattern Recognition (CVPR),
2015, pp.\ 595--604.

\bibitem{parkhi2012cats}
O.~M.~Parkhi, A.~Vedaldi, A.~Zisserman, C.~V.~Jawahar,
Cats and dogs,
in: Proc.\ IEEE Conf.\ on Computer Vision and Pattern Recognition (CVPR),
2012, pp.\ 3498--3505.

\bibitem{vanhorn2018inaturalist}
G.~Van Horn, O.~Mac Aodha, Y.~Song, Y.~Cui, C.~Sun, A.~Shepard, H.~Adam,
P.~Perona, S.~Belongie,
The iNaturalist species classification and detection dataset,
in: Proc.\ IEEE Conf.\ on Computer Vision and Pattern Recognition (CVPR),
2018, pp.\ 8769--8778.

\bibitem{vanhorn2021benchmarking}
G.~Van Horn, E.~Cole, S.~Beery, K.~Wilber, S.~Belongie, O.~Mac Aodha,
Benchmarking representation learning for natural world image collections,
in: Proc.\ IEEE/CVF Conf.\ on Computer Vision and Pattern Recognition
(CVPR), 2021, pp.\ 12884--12893.

\bibitem{russakovsky2015imagenet}
O.~Russakovsky, J.~Deng, H.~Su, J.~Krause, S.~Satheesh, S.~Ma, Z.~Huang,
A.~Karpathy, A.~Khosla, M.~Bernstein, A.~C.~Berg, L.~Fei-Fei,
ImageNet large scale visual recognition challenge,
Int.\ J.\ Comput.\ Vis.\ 115~(3) (2015) 211--252.

\bibitem{he2016deep}
K.~He, X.~Zhang, S.~Ren, J.~Sun,
Deep residual learning for image recognition,
in: Proc.\ IEEE Conf.\ on Computer Vision and Pattern Recognition (CVPR),
2016, pp.\ 770--778.

\bibitem{loshchilov2019decoupled}
I.~Loshchilov, F.~Hutter,
Decoupled weight decay regularization,
in: Int.\ Conf.\ on Learning Representations (ICLR), 2019.

\bibitem{szegedy2016rethinking}
C.~Szegedy, V.~Vanhoucke, S.~Ioffe, J.~Shlens, Z.~Wojna,
Rethinking the Inception architecture for computer vision,
in: Proc.\ IEEE Conf.\ on Computer Vision and Pattern Recognition (CVPR),
2016, pp.\ 2818--2826.

\bibitem{szegedy2017inception}
C.~Szegedy, S.~Ioffe, V.~Vanhoucke, A.~A.~Alemi,
Inception-v4, Inception-ResNet and the impact of residual connections on
learning,
in: Proc.\ AAAI Conf.\ on Artificial Intelligence, 2017.

\bibitem{huang2017densely}
G.~Huang, Z.~Liu, L.~Van Der Maaten, K.~Q.~Weinberger,
Densely connected convolutional networks,
in: Proc.\ IEEE Conf.\ on Computer Vision and Pattern Recognition (CVPR),
2017, pp.\ 4700--4708.

\bibitem{wang2020deep}
J.~Wang, K.~Sun, T.~Cheng, B.~Jiang, C.~Deng, Y.~Zhao, D.~Liu, Y.~Mu,
M.~Tan, X.~Liu, W.~Liu, B.~Xiao,
Deep high-resolution representation learning for visual recognition,
IEEE Trans.\ Pattern Anal.\ Mach.\ Intell.\ 43~(10) (2020) 3349--3364.

\end{thebibliography}
\end{document}